%% file: main.tex
\newcommand{\CLASSINPUTtoptextmargin}{0.75in}%
\newcommand{\CLASSINPUTbottomtextmargin}{1in}%
\newcommand{\CLASSINPUTinnersidemargin}{0.63in}%
\newcommand{\CLASSINPUToutersidemargin}{0.63in}%
\documentclass[conference,10pt,letterpaper]{IEEEtran}%
\usepackage{amsmath}
\usepackage{graphicx}
\usepackage{multirow}
\usepackage[none]{hyphenat}
\usepackage{float}
\usepackage{subfig}
\usepackage{dblfloatfix}
\usepackage{tikz}
\usepackage{t1enc}
\usepackage{times}
\usepackage{cite}
\input{IMS_modify_IEEEtran_18b_CTAN_V7}

\newcommand\copyrighttext{%
  \footnotesize \textcopyright 2025 IEEE.  Personal use of this material is permitted.  Permission from IEEE must be obtained for all other uses, in any current or future media, including reprinting/republishing this material for advertising or promotional purposes, creating new collective works, for resale or redistribution to servers or lists, or reuse of any copyrighted component of this work in other works.}
\newcommand\copyrightnotice{%
\begin{tikzpicture}[remember picture,overlay]
\node[anchor=south,yshift=10pt] at (current page.south) {\fbox{\parbox{\dimexpr\textwidth-\fboxsep-\fboxrule\relax}{\copyrighttext}}};
\end{tikzpicture}%
}

\begin{document}

\raggedbottom
%
%
%
\title{Modeling Josephson traveling-wave parametric amplifiers with electromagnetic and circuit co-simulation}
%
%
%
\IMSthispaperforfinalpublication
\IMSauthor{%
\IMSauthorblockNAME{
Likai Yang\IMSauthorrefmark{\#$\dagger$1},
Jennifer Wang\IMSauthorrefmark{*$\dagger$},
Mohamed Awida Hassan\IMSauthorrefmark{\#}, 
Philip Krantz\IMSauthorrefmark{\#},
Kevin P. O’Brien\IMSauthorrefmark{*2}
}
\\%
\IMSauthorblockAFFIL{
\IMSauthorrefmark{\#}Keysight Technologies, Santa Rosa, USA\\
\IMSauthorrefmark{*}Department of Electrical Engineering and Computer Science, Massachusetts Institute of Technology, Cambridge, USA\\
\IMSauthorrefmark{$\dagger$}These authors contributed equally
}
\\%
\IMSauthorblockEMAIL{
\IMSauthorrefmark{1}likai.yang@keysight.com
\IMSauthorrefmark{2}kpobrien@mit.edu
}
}
%
\maketitle
\copyrightnotice
%
%
%
\begin{abstract}
Optimizing the performance of cryogenic quantum amplifiers is a key step toward achieving high-fidelity and scalable qubit readout. In this work, we present efficient and accurate modeling of a Josephson traveling-wave parametric amplifier (JTWPA) based on electromagnetic (EM) and circuit co-simulation. In contrast to conventional simulation methods where an equivalent lumped-element circuit model of the JTWPA is required, we directly perform full EM analysis of the device to faithfully determine the linear response. The extracted S-parameters are then fed into a nonlinear harmonic balance simulator with Josephson junctions represented as circuit elements. The simulated linear and gain properties are compared with experimental results and show good agreement.
\end{abstract}
%
%

\section{Introduction}
The advancement of quantum information processing with superconducting qubits relies heavily on the faithful amplification and detection of weak microwave signals, as the probe tones for qubit state readout are typically at the few-photon level \cite{wallraff2004strong,heinsoo2018rapid,krantz2019quantum}. Cryogenic parametric amplifiers, usually leveraging the nonlinearity of kinetic inductance \cite{parker2022degenerate} or Josephson junctions (JJs) \cite{yurke1989observation}, serve as advantageous candidates for this purpose, thanks to their near quantum-limited noise performance and straightforward integration with superconducting qubits. They also find various applications in research areas including radio astronomy \cite{chiong2021low}, study of dark matter axions \cite{backes2021quantum}, and spin detection \cite{vine2023situ}. While early works on parametric amplifiers used single or a few nonlinear components coupled to a microwave resonator to enhance the interaction, traveling-wave amplifiers \cite{macklin2015near,malnou2021three} consist of long chains of nonlinear components later emerged as a more favorable configuration due to large bandwidth and high compression power. Today, Josephson traveling-wave parametric amplifiers (JTWPAs) with multi-GHz bandwidth and saturation power over -100\,dBm \cite{qiu2023broadband} are routinely used for superconducting quantum experiments. Therefore, advances in the modeling, design, and fabrication of these amplifiers will further elevate the performance of quantum microwave devices.

The complex nature of JTWPAs, typically composed of thousands of linear and nonlinear components, imposes challenges on accurate modeling and simulation of the device. The typical design methodology for modeling JTWPAs relies on constructing an equivalent lumped-element model of the device for circuit simulation \cite{grimsmo2017squeezing,guarcello2022modeling,peng2022floquet,peng2022x}. A commonly adopted schematic diagram of such circuit models can be found in Ref.~\cite{macklin2015near}, featuring a nonlinear transmission line periodically shunted with LC resonators. However, extracting precise values of these circuit elements could be challenging and may fail to capture various parasitic effects. On the other hand, electromagnetic (EM) modeling of device layout could be pursued, while limited only to linear response. Thus, implementing a combination of EM and nonlinear circuit simulation, a process known as co-simulation, would be highly beneficial.

In this work, we demonstrate a novel co-simulation approach to model a uniform JTWPA device using Keysight Advanced Design Systems (ADS). We first perform an EM analysis of the repeating cells by incorporating junction inductance and capacitance as linear component models. The extracted multi-port S-parameters are then combined with JJ circuit elements and cascaded for nonlinear harmonic balance simulation of the entire device. The linear $S_{21}$ response and gain properties are captured from the simulation and compared with experimentally measured data. A close match between them can be observed, and fine features in device response are also reproduced.

The rest of this paper is structured as follows: in Section II we introduce the layout of the device, together with experimental setups for characterizing the device, which serves as the baseline for benchmarking the modeling approach. Section III focuses on the simulation steps and the comparison with experiments is shown. Finally, Section IV summarizes and discusses the main contributions of this work.


\section{Device layout and experiments}
\begin{figure}[t]
\centering
\includegraphics[width=0.5\textwidth]{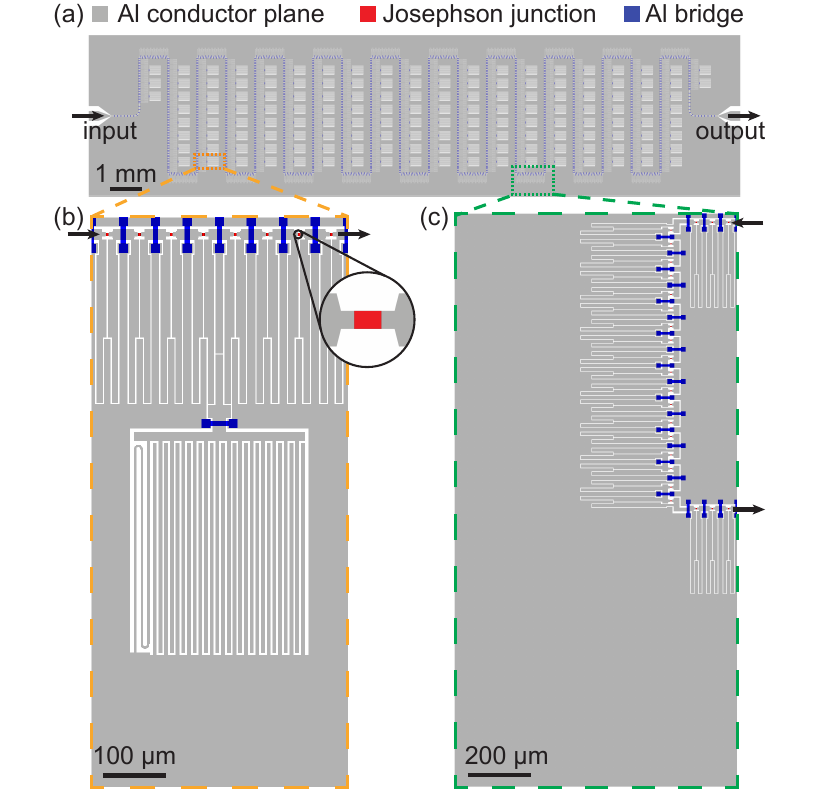}
\caption{JTWPA device layout. (a) The 5x20 mm JTWPA chip consists of a LC ladder transmission line with the nonlinear inductance provided by the JJs, with 1648 unit cells in total. (b) A convenient repeating unit of the JTWPA is the supercell, which includes one phase-matching resonator and 8 unit cells, each including one stub capacitor and three JJs chained in series (represented by a single red block). (c) Meandering this device over the entire chip involves adding a bent section of LC ladder transmission line as supercells at the corners.}
\label{fig:1}
\end{figure}

The device was fabricated from 250\,nm thick superconducting aluminum films on 675\,{\textmu}m thick high-resistivity silicon substrate, using a standard superconducting qubit process. The layout of the device is shown in Fig.~\ref{fig:1}a. The JTWPA consists of a LC ladder transmission line with 1648 unit cells in total, meandered over a 5x20\,mm chip. Each unit cell includes one coplanar capacitor (featuring two closely spaced stubs) to ground, and three identical JJs chained in series, which provide the nonlinear inductance to enable four-wave-mixing parametric amplification. We connect neighboring sections of the ground plane with aluminum metal bridges, supported by silicon dioxide (SiO$_2$) dielectric layer, to minimize unwanted slot-line modes. The height of the metal bridges is set to 750\,nm. For the ease of simulation, we define the "supercell" structure shown in Fig.~\ref{fig:1}b,c. They serve as fundamental units in the simulation, enabling the construction of the entire device by repeating and cascading. As seen in the layout, for every eight unit cells there is a phase-matching resonator that consists of an interdigital capacitor and an inductor (Fig.~\ref{fig:1}b). In order to meander over the entire chip, the supercells at the corners are re-designed as a bent segment of LC ladder transmission line (Fig.~\ref{fig:1}c).

\begin{figure}[t]
\centering
\includegraphics[width=0.5\textwidth]{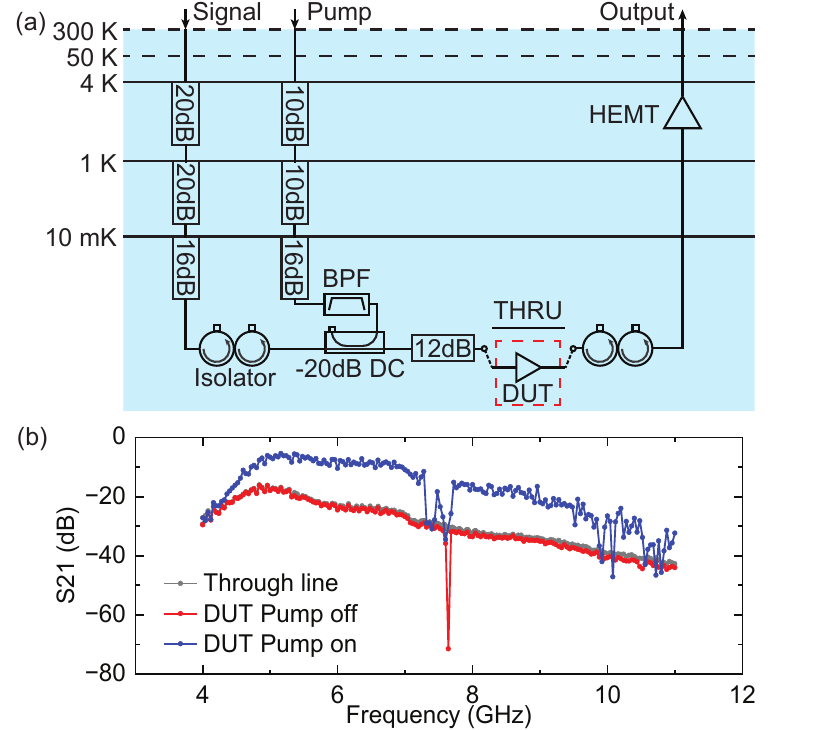}
\caption{Experimental setup for characterizing the JTWPA. (a) Wiring inside the cryostat. The signal and pump lines pass through a series of cryogenic attenuators and are applied to the device under test (DUT). Cryogenic switches are used to switch between DUT and a through line for gain calibration. BPF: 6.85\,GHz to 7.85\,GHz band pass filter, DC: directional coupler, HEMT: high electron mobility transistor. (b) Example data with pump on and off, together with the through line calibration. Gain can be extracted by subtracting the DUT pump on trace and the through line trace. The pump used is 7.5\,GHz and -71.8\,dBm.}
\label{fig:2}
\end{figure}

After fabrication, the JTWPA device was wire bonded to a PCB and mounted in a gold-plated copper package for thermalization. It was then cooled down to below 20\,mK in a dilution refrigerator for characterization. The simplified measurement configuration within the cryostat is sketched in Fig.~\ref{fig:2}a. The probe signal and the parametric pump are sent into the device using two separate lines and combined by a directional coupler which has 20\,dB loss at the coupling port. Cryogenic attenuators with a total of 48\,dB attenuation are added to the pump line to reduce thermal noise. The additional loss from room temperature and cryogenic cables is estimated to be 15\,dB at the pump frequency, which results in a total estimated attenuation of 83\,dB for the pump. The output from the JTWPA is amplified by a high-electron-mobility transistor (HEMT) at the 4\,K stage before digitization with room temperature electronics.

\begin{figure}[b]
\centering
\includegraphics[width=0.5\textwidth]{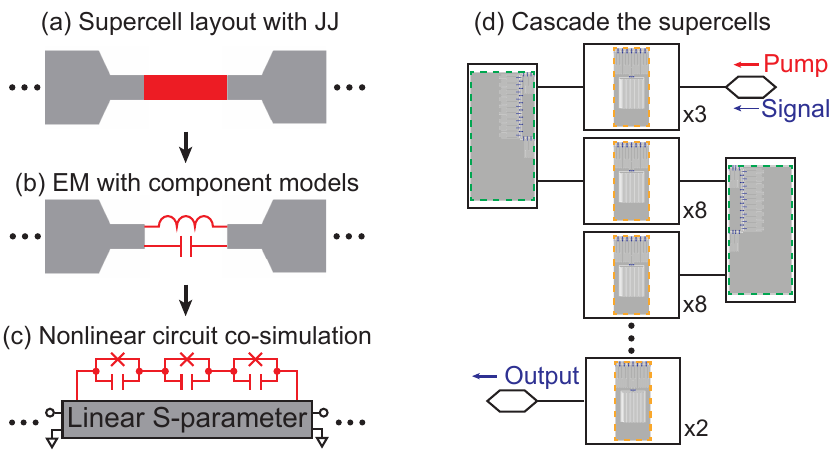}
\caption{Steps to set up the simulation. Step (a): the layout of the supercells from Fig.~\ref{fig:1}b,c is first imported into ADS. Step (b): EM simulation is performed by replacing the JJs with linear capacitance and inductance as component models. Step (c): the S-parameters extracted from EM analysis are incorporated with JJs in nonlinear circuit simulations. Step (d): the supercells are cascaded into the entire device and simulated with harmonic balance simulator.}
\label{fig:3}
\end{figure}

\begin{figure*}[t!]
\centering
\includegraphics[width=0.9\textwidth]{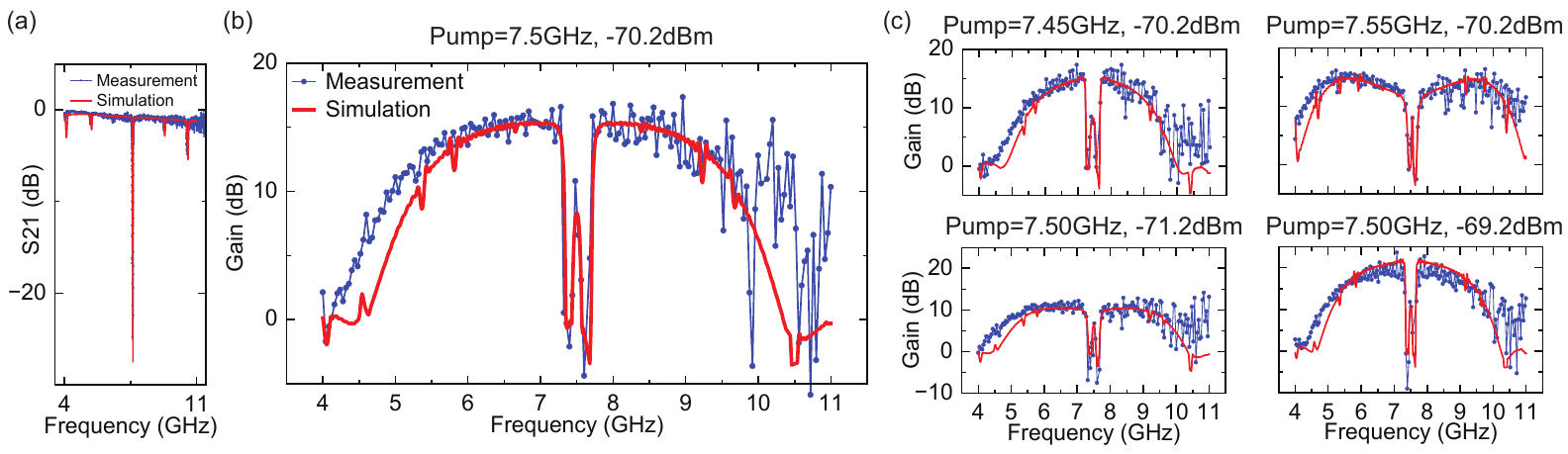}
\caption{Comparison of simulation and experiments. (a) The simulated and measured transmission $S_{21}$ of the device. The measured transmission is normalized to 0\,dB by calibrating out a through line. The resonance frequency as well as side features match well. (b) The gain spectrum simulated with a pump at 7.5\,GHz and -70.2\,dBm. The corresponding experimental gain is measured with 7.5\,GHz and -71.8\,dBm power. (c) Varying pump power by $\pm$1\,dB and frequency by $\pm$50\,MHz in the simulation. The experimental conditions are varied accordingly. Good match is achieved in these gain profiles.}
\label{fig:4}
\end{figure*}

The gain of the device can be extracted by comparing its transmission ($S_{21}$) from the signal to the output port relative to a through line. Typical results of this measurement is plotted in Fig.~\ref{fig:2}b. With the pump turned off, the system exhibits an overall transmission around -30\,dB at 7\,GHz. A sharp dip at 7.62\,GHz can be seen, corresponding to the resonance of the phase-matching resonators. After applying a pump tone at 7.5\,GHz with 11.2\,dBm power, which translates to an estimated power of -71.8\,dBm at the device, gain of the JTWPA device can be observed with an increased $S_{21}$. The gain spectrum can then be calculated by the $S_{21}$ difference relative to the through line, while a systematic mapping of device operation points can be done by sweeping the pump power and frequency.


\section{Simulation}

The steps to set up our simulation is illustrated in Fig.~\ref{fig:3}. We start by importing the layout of the supercells (Fig.~\ref{fig:1}b and c). The JJs are replaced with their linear inductance and capacitance as component models. Full EM analysis is performed on the supercells using a method of moments (MoM) solver. Note that the substrate layers and metal bridges can be accurately represented by defining the corresponding material stack in ADS. On the other hand, due to the limitation of MoM on handling arbitrary dielectric features, the effect of silicon dioxide layer beneath the metal bridges is approximated by reducing the bridge height to get a matching capacitance to the ground. i.e. The height is reduced by a factor of $\epsilon_{SiO_2}=4.6$ to 163\,nm. The EM simulation results are then exported to a circuit simulator as S-parameters. The linear inductors and capacitors are replaced with 3 JJs (embedded circuit component in ADS) in series, which matches the actual device configuration. The critical current and junction capacitance of the JJs are set to $I_c=5.36$\,{\textmu}A and $C_j=201.8$\,fF, respectively, which were characterized in a separate experiment. Finally, the supercells are cascaded into actual device configuration. A harmonic balance simulation is applied by sending a pump and a weak signal tone, and the nonlinear response of the device is extracted. We also note that the mesh used in the EM simulation is set to 200 cells per wavelength (CpW) with edge mesh enabled for the conductors, so that a supercell can be efficiently simulated within 3 minutes, which is a great advantage of using MoM. The gain spectrum with 10\,MHz resolution takes around 15 minutes to simulate. 

The linear response of the device is first simulated and plotted together with experimental results in Fig.~\ref{fig:4}a. We note that due to the excess amount of oxygen incorporated in the aluminum deposition, the kinetic inductance of our film is increased \cite{rotzinger2016aluminium,zhang2019microresonators} and become non-negligible. To account for this effect, we take advantage of the embedded superconductor model in ADS and incorporate a London penetration depth of 195\,nm for the aluminum film. By doing so, a good match in resonator frequency is achieved between simulation and experiments. The simulation also accurately reproduces the multiple side dips seen in the measured $S_{21}$, which is likely due to the impedance mismatch at the corner supercells. The simulated gain profile from the harmonic balance simulator is shown in Fig.~\ref{fig:4}b, with the pump frequency at 7.5\,GHz and a power of -70.2\,dBm. The results align well with the experiments where a pump tone of 7.5\,GHz and -71.8\,dBm is injected into the device. The discrepancy in pump power could be attribute to the measurement error in estimating the exact pump power at the device. The side features in the simulated gain curve can also be attributed to the impedance mismatch effect but have shifted frequencies due to the presence of the strong pump. In the measured data, matched fine features are also visible at lower frequencies but cannot be identified at higher side of the spectrum due to limited signal-to-noise ratio. Small ripples can also be seen in the overall gain spectrum likely because the impedance of the transmission line deviates slightly from 50\,$\Omega$. The free spectral range (FSR) of these oscillations matches well with a weak cavity formed by the electrical length of the device.

To further validate the accuracy of our modeling method, we sweep the pump power and frequency in the simulation and compare with experimental data. The results are demonstrated in Fig.~\ref{fig:4}c. With the pump frequency detuned by $\pm$50\,MHz and the pump power varied by $\pm$1\,dB, the change in gain spectra follows closely with experiments, suggesting that our simulation is self-consistent. Such sweeping simulation would also be very helpful for optimizing device performance under certain pump conditions.


\section{Conclusion and discussion}
In conclusion, we have developed a co-simulation method for modeling JTWPAs, by combining linear EM analysis of the device layout and nonlinear circuit simulation of JJ circuit elements. Using this approach, a uniform JTWPA device is simulated and the results are compared with experiments. A reasonable match between the two is achieved. Fine features in device response due to impedance mismatch are well-captured in simulation. We note that while the frequencies of these features align with the experiments, there is some deviation in their amplitude. This could be due to other effects that are not included in the simulation, such as parasitic coupling between adjacent supercells and variability in JJ properties. The deviation of the junction capacitance may also contribute to the gain bandwidth discrepancy between simulation and measurement. These effects could be optimized in future simulations, so that a better representation of the actual device can be reached. Compared to pure circuit simulation from lumped-element models, our approach skips the step of finding the equivalent values of each components, thus would be more convenient to implement and more accurate in reproducing the device response. It can also be easily generalized to different device configurations and other nonlinear devices such as kinetic inductance parametric amplifiers. In light of these advantages, our method represents a valuable tool for guiding the design of future generations of quantum parametric amplifiers.


\section*{Acknowledgment}

\newcommand{\IMSacktext}{%
This work was funded in part by the AWS Center for Quantum Computing and by the MIT Research Support Committee from the NEC Corporation Fund for Research in Computers and Communications. This work was carried out in part through the use of MIT.nano's facilities. J. Wang acknowledges support from the NSF GRFP. The authors acknowledge the Keysight EDA development team for their contribution in enabling the JTWPA workflow in Keysight ADS.
}
\IMSdisplayacksection{\IMSacktext}


\bibliographystyle{IEEEtran}

\bibliography{IEEEabrv,reference}

\end{document}

%% file: IMS_modify_IEEEtran_18b_CTAN_V7.tex
\makeatletter

\def\@IMSauthorblockNAMEstyle{\normalfont\IMSauthorsize}
\def\@IMSauthorblockAFFILstyle{\normalfont\IMSaffilsize}
\def\@IMSauthorblockEMAILstyle{\normalfont\IMSaffilsize}
\def\IMSauthorblockNAME#1{%
\relax\@IMSauthorblockNAMEstyle%
#1%
}%
\def\IMSauthorblockAFFIL#1{%
\relax\@IMSauthorblockAFFILstyle%
\vskip\@IEEEauthorblockAtopspace
#1%
}%
\def\IMSauthorblockEMAIL#1{%
\relax\@IMSauthorblockEMAILstyle%
\vskip\@IEEEauthorblockAtopspace
#1%
}%
\newcommand{\IMSauthor}[1]{%
\ifIsBlindReviewVersion%
\author{\phantom{\parbox{\textwidth}{\center\relax#1}}}%
\else%
\author{\parbox{\textwidth}{\center\relax#1}}%
\fi%
}%
\newif\ifIsBlindReviewVersion

\def\IMSthispaperforfinalpublication{\IsBlindReviewVersionfalse}
\def\@maketitle{\newpage
\bgroup\par\addvspace{0.5\baselineskip}\centering%
\ifCLASSOPTIONtechnote
   {\bfseries\large\@IEEEcompsoconly{\sffamily}\@title\par}\vskip 1.3em{\lineskip .5em\@IEEEcompsoconly{\sffamily}\@author
   \@IEEEspecialpapernotice\par{\@IEEEcompsoconly{\vskip 1.5em\relax
   \@IEEEtitleabstractindextextbox{\@IEEEtitleabstractindextext}\par
   \hfill\@IEEEcompsocdiamondline\hfill\hbox{}\par}}}\relax
\else
   \vskip0.2em{\IMStitlesize\ifCLASSOPTIONtransmag\bfseries\LARGE\fi\@IEEEcompsoconly{\sffamily}\@IEEEcompsocconfonly{\normalfont\normalsize\vskip 2\@IEEEnormalsizeunitybaselineskip
   \bfseries\Large}\@title\par}\vskip1.0em\par
   \ifCLASSOPTIONconference%
      {\@IEEEspecialpapernotice\mbox{}\vskip\@IEEEauthorblockconfadjspace%
       \mbox{}\hfill\begin{@IEEEauthorhalign}\@author\end{@IEEEauthorhalign}\hfill\mbox{}\par}\relax
   \else
      \ifCLASSOPTIONpeerreviewca
         {\@IEEEcompsoconly{\sffamily}\@IEEEspecialpapernotice\mbox{}\vskip\@IEEEauthorblockconfadjspace%
          \mbox{}\hfill\begin{@IEEEauthorhalign}\@author\end{@IEEEauthorhalign}\hfill\mbox{}\par
          {\@IEEEcompsoconly{\vskip 1.5em\relax
           \@IEEEtitleabstractindextextbox{\@IEEEtitleabstractindextext}\par\hfill
           \@IEEEcompsocdiamondline\hfill\hbox{}\par}}}\relax
      \else
         \ifCLASSOPTIONtransmag
           {\@IEEEspecialpapernotice\mbox{}\vskip\@IEEEauthorblockconfadjspace%
            \mbox{}\hfill\begin{@IEEEauthorhalign}\@author\end{@IEEEauthorhalign}\hfill\mbox{}\par
           {\vspace{0.5\baselineskip}\relax\@IEEEtitleabstractindextextbox{\@IEEEtitleabstractindextext}\vspace{-1\baselineskip}\par}}\relax
         \else
           {\lineskip.5em\@IEEEcompsoconly{\sffamily}\sublargesize\@author\@IEEEspecialpapernotice\par
           {\@IEEEcompsoconly{\vskip 1.5em\relax
            \@IEEEtitleabstractindextextbox{\@IEEEtitleabstractindextext}\par\hfill
            \@IEEEcompsocdiamondline\hfill\hbox{}\par}}}\relax
         \fi
      \fi
   \fi
\fi\par\addvspace{0.0\baselineskip}\egroup}

\def\IMStitlesize{\@setfontsize{\IMStitlesize}{18}{21pt}}
\def\IMSauthorsize{\@setfontsize{\IMSauthorsize}{12}{13pt}}
\def\IMSaffilsize{\@setfontsize{\IMSaffilsize}{12}{13pt}}
\def\IMScaptionsize{\@setfontsize{\IMScaptionsize}{8}{9pt}}
\def\IMSbibsize{\@setfontsize{\IMSbibsize}{8}{9pt}}

\def\@IEEEauthorblockNstyle{\IMSauthorsize\@IEEEcompsocnotconfonly{\sffamily}\@IEEEcompsocconfonly{\large}}
\def\@IEEEauthorblockAstyle{\IMSaffilsize\@IEEEcompsocnotconfonly{\sffamily}\@IEEEcompsocconfonly{\itshape}\@IEEEcompsocconfonly{\large}}
\def\@IEEEauthordefaulttextstyle{\IMSauthorsize\@IEEEcompsocnotconfonly{\sffamily}\sublargesize}

\def\thebibliography#1{\section*{\refname}%
    \addcontentsline{toc}{section}{\refname}%
    \IMSbibsize\@IEEEcompsocconfonly{\small}\vskip 0.3\baselineskip plus 0.1\baselineskip minus 0.1\baselineskip
    \list{\@biblabel{\@arabic\c@enumiv}}%
    {\settowidth\labelwidth{\@biblabel{#1}}%
    \leftmargin\labelwidth
    \advance\leftmargin\labelsep\relax
    \itemsep \IEEEbibitemsep\relax
    \usecounter{enumiv}%
    \let\p@enumiv\@empty
    \renewcommand\theenumiv{\@arabic\c@enumiv}}%
    \let\@IEEElatexbibitem\bibitem%
    \def\bibitem{\@IEEEbibitemprefix\@IEEElatexbibitem}%
\def\newblock{\hskip .11em plus .33em minus .07em}%
\ifCLASSOPTIONtechnote\sloppy\clubpenalty4000\widowpenalty4000\interlinepenalty100%
\else\sloppy\clubpenalty4000\widowpenalty4000\interlinepenalty500\fi%
    \sfcode`\.=1000\relax}

\long\def\@makecaption#1#2{%
\ifx\@captype\@IEEEtablestring%
\par\@IEEEtabletopskipstrut
\else
\@IEEEfigurecaptionsepspace
\fi
\setbox\@tempboxa\hbox{\normalfont\IMScaptionsize {#1.}\nobreakspace\nobreakspace #2}%
\ifdim \wd\@tempboxa >\hsize%
\setbox\@tempboxa\hbox{\normalfont\IMScaptionsize {#1.}\nobreakspace\nobreakspace}%
\parbox[t]{\hsize}{\normalfont\IMScaptionsize\noindent\unhbox\@tempboxa#2}%
\else
\ifCLASSOPTIONconference \hbox to\hsize{\normalfont\IMScaptionsize\hfil\box\@tempboxa\hfil}%
\else \hbox to\hsize{\normalfont\IMScaptionsize\box\@tempboxa\hfil}%
\fi\fi
\ifx\@captype\@IEEEtablestring%
\@IEEEtablecaptionsepspace
\else
\fi}

\newlength\tablecaptiontotableskip
\newlength\figuretocaptionskip
\def\@IEEEfigurecaptionsepspace{\vskip\figuretocaptionskip\relax}%
\def\@IEEEtablecaptionsepspace{\vskip\tablecaptiontotableskip\relax}%

\def\abstract{\normalfont%
\@IEEEabskeysecsize\bfseries\textit{\abstractname}\,\bfseries\textit{---}\,%
\@IEEEgobbleleadPARNLSP}%

\def\IEEEkeywords{\normalfont%
\@IEEEabskeysecsize\bfseries\textit{\IEEEkeywordsname}\,\bfseries\textit{---}\,%
\@IEEEgobbleleadPARNLSP}%
\def\endIEEEkeywords{\relax\vspace{0.67ex}%
\par\if@twocolumn\else\endquotation\fi%
\normalsize\normalfont}%

\DeclareRobustCommand*{\IMSauthorrefmark}[1]{\raisebox{0pt}[0pt][0pt]{\textsuperscript{\footnotesize{#1}}}}%
\def\@IEEEauthorblockNtopspace{0ex}
\def\@IEEEauthorblockAtopspace{1mm}
\def\tablename{Table}
\def\thetable{\arabic{table}}
\def\IEEEkeywordsname{Keywords}
\def\subsubsection{\@startsection{subsubsection}{3}{\z@}{1.5ex plus 1.5ex minus 0.5ex}%
{0.7ex plus .5ex minus 0ex}{\normalfont\normalsize\itshape}}%
\def\@seccntformat#1{\csname the#1dis\endcsname\relax}
\def\thesectiondis{\thesection.\hskip 0.5em}
\def\thesubsectiondis{{\hbox to\parindent{\Alph{subsection}.}}}
\def\thesubsubsectiondis{{\hbox to \parindent{\arabic{subsubsection})}}}
\def\theparagraphdis{{\hbox to \parindent{\alph{paragraph})}}}
\IEEEilabelindentA \parindent
\IEEEilabelindent \IEEEilabelindentA
\IEEEelabelindent \parindent
\IEEEdlabelindent \parindent
\IEEElabelindent \parindent

\newlength\@IMSparindent
\newcommand\IMSdisplayacksection[1]{%
\ifIsBlindReviewVersion%
\noindent\phantom{\parbox[t]{\columnwidth}{\normalbaselines\setlength{\parindent}{\@IMSparindent}{#1}\strut}}
\else%
\noindent\parbox[t]{\columnwidth}{\normalbaselines\setlength{\parindent}{\@IMSparindent}{#1}\strut}%
\fi%
}%

\makeatother

%% file: main.bbl
\begin{thebibliography}{10}
\providecommand{\url}[1]{#1}
\csname url@samestyle\endcsname
\providecommand{\newblock}{\relax}
\providecommand{\bibinfo}[2]{#2}
\providecommand{\BIBentrySTDinterwordspacing}{\spaceskip=0pt\relax}
\providecommand{\BIBentryALTinterwordstretchfactor}{4}
\providecommand{\BIBentryALTinterwordspacing}{\spaceskip=\fontdimen2\font plus
\BIBentryALTinterwordstretchfactor\fontdimen3\font minus \fontdimen4\font\relax}
\providecommand{\BIBforeignlanguage}[2]{{%
\expandafter\ifx\csname l@#1\endcsname\relax
\typeout{** WARNING: IEEEtran.bst: No hyphenation pattern has been}%
\typeout{** loaded for the language `#1'. Using the pattern for}%
\typeout{** the default language instead.}%
\else
\language=\csname l@#1\endcsname
\fi
#2}}
\providecommand{\BIBdecl}{\relax}
\BIBdecl

\bibitem{wallraff2004strong}
A.~Wallraff, D.~I. Schuster, A.~Blais, L.~Frunzio, R.-S. Huang, J.~Majer, S.~Kumar, S.~M. Girvin, and R.~J. Schoelkopf, ``Strong coupling of a single photon to a superconducting qubit using circuit quantum electrodynamics,'' \emph{Nature}, vol. 431, no. 7005, pp. 162--167, 2004.

\bibitem{heinsoo2018rapid}
J.~Heinsoo, C.~K. Andersen, A.~Remm, S.~Krinner, T.~Walter, Y.~Salath{\'e}, S.~Gasparinetti, J.-C. Besse, A.~Poto{\v{c}}nik, A.~Wallraff \emph{et~al.}, ``Rapid high-fidelity multiplexed readout of superconducting qubits,'' \emph{Physical Review Applied}, vol.~10, no.~3, p. 034040, 2018.

\bibitem{krantz2019quantum}
P.~Krantz, M.~Kjaergaard, F.~Yan, T.~P. Orlando, S.~Gustavsson, and W.~D. Oliver, ``A quantum engineer's guide to superconducting qubits,'' \emph{Applied physics reviews}, vol.~6, no.~2, 2019.

\bibitem{parker2022degenerate}
D.~J. Parker, M.~Savytskyi, W.~Vine, A.~Laucht, T.~Duty, A.~Morello, A.~L. Grimsmo, and J.~J. Pla, ``Degenerate parametric amplification via three-wave mixing using kinetic inductance,'' \emph{Physical Review Applied}, vol.~17, no.~3, p. 034064, 2022.

\bibitem{yurke1989observation}
B.~Yurke, L.~Corruccini, P.~Kaminsky, L.~Rupp, A.~Smith, A.~Silver, R.~Simon, and E.~Whittaker, ``Observation of parametric amplification and deamplification in a josephson parametric amplifier,'' \emph{Physical Review A}, vol.~39, no.~5, p. 2519, 1989.

\bibitem{chiong2021low}
C.-C. Chiong, Y.~Wang, K.-C. Chang, and H.~Wang, ``Low-noise amplifier for next-generation radio astronomy telescopes: Review of the state-of-the-art cryogenic lnas in the most challenging applications,'' \emph{IEEE Microwave Magazine}, vol.~23, no.~1, pp. 31--47, 2021.

\bibitem{backes2021quantum}
K.~M. Backes, D.~A. Palken, S.~A. Kenany, B.~M. Brubaker, S.~Cahn, A.~Droster, G.~C. Hilton, S.~Ghosh, H.~Jackson, S.~K. Lamoreaux \emph{et~al.}, ``A quantum enhanced search for dark matter axions,'' \emph{Nature}, vol. 590, no. 7845, pp. 238--242, 2021.

\bibitem{vine2023situ}
W.~Vine, M.~Savytskyi, A.~Vaartjes, A.~Kringh{\o}j, D.~Parker, J.~Slack-Smith, T.~Schenkel, K.~M{\o}lmer, J.~C. McCallum, B.~C. Johnson \emph{et~al.}, ``In situ amplification of spin echoes within a kinetic inductance parametric amplifier,'' \emph{Science advances}, vol.~9, no.~10, p. eadg1593, 2023.

\bibitem{macklin2015near}
C.~Macklin, K.~O’brien, D.~Hover, M.~Schwartz, V.~Bolkhovsky, X.~Zhang, W.~Oliver, and I.~Siddiqi, ``A near--quantum-limited josephson traveling-wave parametric amplifier,'' \emph{Science}, vol. 350, no. 6258, pp. 307--310, 2015.

\bibitem{malnou2021three}
M.~Malnou, M.~Vissers, J.~Wheeler, J.~Aumentado, J.~Hubmayr, J.~Ullom, and J.~Gao, ``Three-wave mixing kinetic inductance traveling-wave amplifier with near-quantum-limited noise performance,'' \emph{PRX Quantum}, vol.~2, no.~1, p. 010302, 2021.

\bibitem{qiu2023broadband}
J.~Y. Qiu, A.~Grimsmo, K.~Peng, B.~Kannan, B.~Lienhard, Y.~Sung, P.~Krantz, V.~Bolkhovsky, G.~Calusine, D.~Kim \emph{et~al.}, ``Broadband squeezed microwaves and amplification with a josephson travelling-wave parametric amplifier,'' \emph{Nature Physics}, vol.~19, no.~5, pp. 706--713, 2023.

\bibitem{grimsmo2017squeezing}
A.~L. Grimsmo and A.~Blais, ``Squeezing and quantum state engineering with josephson travelling wave amplifiers,'' \emph{npj Quantum Information}, vol.~3, no.~1, p.~20, 2017.

\bibitem{guarcello2022modeling}
C.~Guarcello, G.~Avallone, C.~Barone, M.~Borghesi, S.~Capelli, G.~Carapella, A.~P. Caricato, I.~Carusotto, A.~Cian, D.~Di~Gioacchino \emph{et~al.}, ``Modeling of josephson traveling wave parametric amplifiers,'' \emph{IEEE Transactions on Applied Superconductivity}, vol.~33, no.~1, pp. 1--7, 2022.

\bibitem{peng2022floquet}
K.~Peng, M.~Naghiloo, J.~Wang, G.~D. Cunningham, Y.~Ye, and K.~P. O’Brien, ``Floquet-mode traveling-wave parametric amplifiers,'' \emph{PRX Quantum}, vol.~3, no.~2, p. 020306, 2022.

\bibitem{peng2022x}
K.~Peng, R.~Poore, P.~Krantz, D.~E. Root, and K.~P. O’Brien, ``X-parameter based design and simulation of josephson traveling-wave parametric amplifiers for quantum computing applications,'' in \emph{2022 IEEE International Conference on Quantum Computing and Engineering (QCE)}.\hskip 1em plus 0.5em minus 0.4em\relax IEEE, 2022, pp. 331--340.

\bibitem{rotzinger2016aluminium}
H.~Rotzinger, S.~Skacel, M.~Pfirrmann, J.~Voss, J.~M{\"u}nzberg, S.~Probst, P.~Bushev, M.~Weides, A.~Ustinov, and J.~Mooij, ``Aluminium-oxide wires for superconducting high kinetic inductance circuits,'' \emph{Superconductor Science and Technology}, vol.~30, no.~2, p. 025002, 2016.

\bibitem{zhang2019microresonators}
W.~Zhang, K.~Kalashnikov, W.-S. Lu, P.~Kamenov, T.~DiNapoli, and M.~Gershenson, ``Microresonators fabricated from high-kinetic-inductance aluminum films,'' \emph{Physical Review Applied}, vol.~11, no.~1, p. 011003, 2019.

\end{thebibliography}
